\documentclass[lineno]{jfm}
\usepackage{graphicx}
\usepackage{newtxtext}
\usepackage{newtxmath}
\usepackage{natbib}
\usepackage{hyperref}
\usepackage{printlen}
\usepackage{float}
\usepackage{inputenc}
\usepackage[capitalise]{cleveref}
\hypersetup{
colorlinks = true,
urlcolor   = blue,
citecolor  = black,
}

\newcommand{\RomanNumeralCaps}[1]
\linenumbers
\newcommand{\grad}{\boldsymbol{\nabla}}
\newcommand{\wo}{\mathrm{Wo}}
\newcommand{\vf}{\boldsymbol{v}}

\title{Unsteady drag force on an immersed sphere oscillating near a wall}
\author{Zaicheng Zhang\aff{1},
{Vincent Bertin\aff{2}},
Martin H. Essink\aff{2},
Hao Zhang\aff{1},
Nicolas Fares\aff{1},
Zaiyi Shen\aff{3},
Thomas Bickel\aff{1},
Thomas Salez\aff{1}\corresp{\email{thomas.salez@cnrs.fr}}
\and {Abdelhamid Maali\aff{1}
\corresp{\email{abdelhamid.maali@u-bordeaux.fr}}}
}
\affiliation{\aff{1}Univ. Bordeaux, CNRS, LOMA, UMR 5798, 33405 Talence, France.
\aff{2}Physics of Fluids Group, Faculty of Science and Technology, Mesa+ Institute, University of Twente, 7500 AE Enschede, The Netherlands.
\aff{3}State Key Laboratory for Turbulence and Complex Systems, Department of Mechanics and Engineering Science,
College of Engineering, Peking University, Beijing 100871, China}

\begin{document}
\maketitle

\begin{abstract}
The unsteady hydrodynamic drag exerted on an oscillating sphere near a planar wall is addressed experimentally,  theoretically, and numerically. The experiments are performed by using colloidal-probe Atomic Force Microscopy (AFM) in thermal noise mode. The natural resonance frequencies and quality factors are extracted from the measurement of the power spectrum density of the probe oscillation for a broad range of gap distances and Womersley numbers. 
The shift in the natural resonance frequency of the colloidal probe as the probe goes close to a solid wall infers the wall-induced variations of the effective mass of the probe. Interestingly, a crossover from a positive to a negative shift is observed as the Womersley number increases. In order to rationalize the results, the confined unsteady Stokes equation is solved numerically using a finite-element method, as well as asymptotic calculations.
The in-phase and out-of-phase terms of the hydrodynamic drag acting on the sphere are obtained and agree well to the  experimental results. All together, the experimental,  theoretical, and numerical results show that the hydrodynamic force felt by an immersed sphere oscillating near a wall is highly dependent on the Womersley number.
\end{abstract}


\section{Introduction}
\label{sec:intro}
The motion of particles in a fluid is one of the central problems in fluid mechanics, across many scales. The hydrodynamic drag force exerted by the fluid on the particles is the fundamental quantity that dictates the motion. Applications include the sedimentation of synthetic entities, the swimming of biological microorganisms~\citep[see][]{wang2012unsteady,wei2019zero,wei2021measurements,redaelli2022hydrodynamic}, blood flows~\citep[see][]{ku1997blood}, peristaltic pumping~\citep[see][]{shapiro1969peristaltic}, microfluidic flows~\citep[see][]{dincau2020pulsatile}, Brownian motion at short times~\citep[see][]{felderhof2005effect,mo2019highly}, etc... 
At small Reynolds number, while the steady, bulk, Stokes’ drag force exerted on a translating sphere is well known, addressing further the transient contributions is more intricate -- even though the implications of such effects are potentially numerous. 

For an isolated spherical particle with radius $R$ translating in a viscous liquid at velocity $\boldsymbol{V}$, the bulk drag force $\boldsymbol{F}$ at small Reynolds number is given by the Basset-Boussinesq-Oseen (BBO) expression~\citep[see][]{basset1888treatise,maxey1983equation,lovalenti1993hydrodynamic,landau1987fluid}: 
\begin{equation}
\label{eq:stokes_isolatedsphere}
\boldsymbol{F}= -6\pi\eta R\boldsymbol{V} - 6R^2\sqrt{\pi\rho \eta}\int_{-\infty}^t \frac{1}{\sqrt{t-\tau}} \frac{\mathrm{d} \boldsymbol{V}}{\mathrm{d}\tau}\mathrm{d}\tau - \frac{2\pi \rho R^3}{3}\frac{\mathrm{d}\boldsymbol{V}}{\mathrm{d}t},
\end{equation}
where $\rho$ and $\eta$ are the density and dynamic viscosity of the viscous liquid, respectively.
The right-hand side of the latter equation includes three terms successively: a Stokes viscous force, a Basset memory term, and an added-mass term. The Basset force originates from the diffusive nature of vorticity within the unsteady Stokes equation, and the added-mass force can be interpreted as an inertial effect due to the displaced fluid mass. Equation~\eqref{eq:stokes_isolatedsphere} provides a good description of particle dynamics in a large variety of particle-laden and multi-phase flows, as long as the particle Reynolds number is small~\citep[see][]{balachandar2010turbulent}. 

Nevertheless, the effect of nearby solid boundaries on the unsteady drag is still an open question. The canonical situation is that of an immersed sphere oscillating near a planar rigid surface. Some asymptotic expressions of the drag in the large-distance limit have been derived recently, by using a point-particle approximation together with the method of images~\citep{felderhof2005effect,felderhof2012hydrodynamic,simha2018unsteady}, or by using low or high frequency expansions of the unsteady Stokes equations~\citep{fouxon2018fundamental}. However, theoretical descriptions of the confined limit, \textit{i.e.} where the sphere is in close proximity to the surface, are scarce. We thus aim here at investigating the unsteady drag, in the full spatial range from bulk to confinement, by combining numerical simulations, asymptotic calculations and colloidal-probe Atomic Force Microscopy (AFM) experiments. 

AFM colloidal-probe methods and their Surface Force Apparatus (SFA) analogues, have been first introduced in the 1990's in order to measure molecular interactions (\textit{e.g.} electrostatic, van der Waals, ...) between surfaces~\citep[see][]{butt1991measuring,ducker1991direct,butt2005force}. Recently, these methods have been extended and used to study flow under micro-to-nanometric confinement, \textit{e.g.} near soft~\citep[see][]{leroy2011hydrodynamic,leroy2012hydrodynamic,villey2013effect,guan2017noncontact, zhang2022contactless} or capillary interfaces~\citep[see][]{manor2008hydrodynamic,vakarelski2010dynamic,manica2016impact,maali2017viscoelastic,wang2018viscocapillary,bertin2021contactless},
using complex fluids~\citep[see][]{comtet2017pairwise,comtet2017nanoscale,comtet2019atomic}, or to measure the friction at solid-liquid interfaces~\citep[see][]{cottin2003low,maali2008measurement,cross2018wall}, and electrohydrodynamic effects~\citep[see][]{liu2018electroviscous, liu2015amplitude,zhao2020electroviscous, Matus2022electroviscous}, etc... More specifically, for dynamic colloidal AFM measurements, a micron-size spherical colloidal probe is placed in a viscous fluid, in the vicinity of a surface, with a probe-surface distance $D$. Then, the probe is driven to oscillate without direct contact, via either acoustic excitation or thermal noise. The force exerted on the sphere is inferred from the colloidal motion, trough the cantilever's deflection, which allows to extract specific information on the confined surfaces or fluid properties. We point out that other experimental techniques were used to probe the bulk streaming flow around an oscillating sphere at finite Reynolds numbers, like particle visualization techniques~\citep{kotas2007visualization,otto2008measurements}, and optical tweezers~\citep{bruot2021direct}. 

If the typical angular frequency of the flow is $\omega$, then the vorticity diffuses on a typical distance $\delta \sim \sqrt{\eta/(\rho\omega)}$ called the viscous penetration length. The dynamic force measurements are usually restricted to low Reynolds numbers, low probing frequencies, and to the confined regime where  $D\ll R$. In such a case, the penetration length is large, the flow is mainly located in the confined fluid layer, it is purely viscous and quasi-steady, and the lubrication theory holds~\citep[see][]{reynolds1886iv, leroy2011hydrodynamic}. Consequently, in all the above examples, the fluid inertial effects are disregarded in the analysis of the measured hydrodynamic force. However, when the colloidal probe oscillates at high frequencies, the penetration depth $\delta$ becomes smaller and comparable to the characteristic length scale of the flow. Thus, unsteady effects become important~\citep[see][]{clarke2005drag}. The relevant dimensionless number to characterize the crossover to such a regime is the Womersley number $\wo = R \sqrt{\omega / \nu}$, the square of which corresponds to the ratio between the typical diffusion time scale $R^2/\nu$, and the period of the oscillation. Inertial effects should be predominant when it takes more time for the velocity field to diffuse than for the sphere to oscillate, \textit{i.e.} $\wo > 1$. In such a situation, the hydrodynamic force exerted
on the sphere is not only a viscous lubrication drag, but it also contains contributions due to the fluid inertia, which were partly studied in previous works~\citep[see][]{sader1998frequency,benmouna2002hydrodynamic,clarke2005drag,devailly2020long}. 

The article is organized as follows. In Sec. \ref{sec:exp}, we introduce the experimental method of thermal noise AFM and present the typical experimental results.
We show that, as the distance to the wall is reduced, the natural frequency  increases for low Womersley numbers but decreases for high Womersley numbers. In contrast, the dissipation monotonically increases with decreasing distance for all Womersley numbers. In order to rationalize the results, in Sec. \ref{sec:model}, we compute the hydrodynamic drag force in terms of added mass and dissipation, in the asymptotic limit of large distance, and we perform a detailed calculation in the Low-Womersley limit using the Lorentz reciprocal theorem. Furthermore, a finite-element method is employed to obtain the full numerical solution in all regimes. Finally, the experimental, theoretical and numerical results are summarized and compared in Sec.~\ref{sec:results}. Mainly, the variation of the resonance frequency is rationalized by the change of the effective mass with distance and Womersley number. 

\section{Experiments}
\label{sec:exp}
\subsection{Colloidal-probe AFM setup}
\begin{figure}
\centering 
\includegraphics{{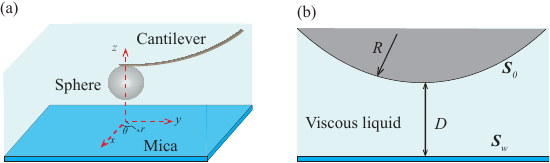}}
\caption{\label{fig:fig1} 
Schematics of the system. A borosilicate sphere with radius $R$ is glued at the end of an AFM cantilever, and thermally fluctuates within a viscous liquid and near a mica substrate, with a distance $D$ between the sphere and the substrate. The sphere and the mica surfaces are denoted as $\mathcal{S}_0$ and $\mathcal{S}_{\textrm{w}}$, respectively.}
\end{figure}

A schematic of the experimental system is shown in~\cref{fig:fig1}(a). A borosilicate sphere (MOSci Corporation, radius $R=27 \pm 0.5 \,\mathrm{\mu m}$) is glued (Epoxy glue, Araldite) to the end of an AFM cantilever (SNL-10, Brukerprobes), and located near a planar mica surface. The cantilever stiffness $k_\mathrm{c}=0.68\pm 0.05 \,\mathrm{N/m}$ is calibrated using the drainage method proposed by~\cite{craig2001situ}. The experiments were performed using an AFM (Bruker, Dimension3100) in three different liquids, \textit{i.e} water, dodecane and silicone oil, whose densities and dynamic viscosities are $1000$ kg/m$^3$, 1 $\mathrm{mPa\cdot s}$, $750$ kg/m$^3$, 1.34 $\mathrm{mPa\cdot s}$ and $930$ kg/m$^3$, 9.3 $\mathrm{mPa\cdot s}$, respectively, at room temperature. The probe-surface distance $D$ was controlled by an integrated stage step motor. Each separation distance was adjusted by displacing the cantilever vertically using the step motor with precision in position $<0.1 \,\mathrm{\mu m}$. The probe's deflection was directly acquired using an analog to digital (A/D) acquisition card (PCI-4462, NI, USA) with a sample frequency of $200\, \mathrm{kHz}$. The vertical position of the probe was observed to fluctuate due to thermal noise, as discussed in the following section. The amplitude of the sphere's fluctuation remains smaller than $\sim1~\mathrm{nm}$ in all the experiments.

\subsection{Confined thermal dynamics}
\label{subsec:probe-dynamics}
The time-dependent position of the probe is denoted $Z(t)$. We suppose that the probe dynamics can be modelled by a forced harmonic oscillator, as:
\begin{equation}
\label{eq:motion-equation}
m_\infty \ddot{Z}+ \gamma_\infty \dot{Z} + k_\mathrm{c} Z = F_\mathrm{th} + F_\mathrm{int},
\end{equation}
where $m_\infty$ is the effective mass of the probe in the bulk, and where $\gamma_\infty$ is the bulk damping coefficient. These two coefficients correspond to the free dynamics of the probe far from the surface, and can thus be obtained by measuring the resonance properties of the AFM probe in the far field, as shown below. Besides the elastic restoring force by the cantilever of stiffness $k_\mathrm{c}$, and in the absence of conservative forces (\textit{e.g.} van der Waals or electrostatic forces), the two main forces acting on the sphere along the $z$ direction are the random thermal force $F_\mathrm{th}$ and the hydrodynamic interaction force with the wall $F_\mathrm{int}$. The latter corresponds to the deviation of the hydrodynamic drag with respect to the bulk drag force. 

Taking the Fourier transform of~\cref{eq:motion-equation}, we find:
\begin{equation}
\label{eq:motion-equation_fourier}
- m_\infty \omega^2 \tilde{Z} + i\omega \gamma_\infty \tilde{Z} + k_\mathrm{c} \tilde{Z} = \tilde{F}_\mathrm{th} + \tilde{F}_\mathrm{int},
\end{equation}
where $\tilde{f}(\omega) = \frac{1}{2\pi}\int_{-\infty}^{\infty} \, \mathrm{d}t \, f(t) e^{-i \omega t}$ is the Fourier transform of the function $f(t)$. The real and imaginary parts of $\tilde{F}_\mathrm{int}$ correspond to an inertial force and a dissipative force, respectively, that can be recast into:
\begin{equation}
\label{eq:interaction-force_fourier}
    \tilde{F}_\mathrm{int} = m_\mathrm{int} \omega^2 \tilde{Z} - i\omega\gamma_\mathrm{int} \tilde{Z},
\end{equation}
where $m_\mathrm{int}$ and $\gamma_\mathrm{int}$ are the wall-induced variations of the effective mass and dissipation coefficient. For the sake of simplicity, we neglect in the following the possible frequency dependencies of $m_\mathrm{int}$ and $\gamma_\mathrm{int}$. With this assumption, and injecting \cref{eq:interaction-force_fourier} into \cref{eq:motion-equation_fourier}, the probe's motion follows a thermally-forced harmonic oscillator dynamics with a spring constant $ k_\mathrm{c}$, an effective damping coefficient $\gamma \equiv \gamma_\infty + \gamma_\mathrm{int}$ and an effective mass $m \equiv m_\infty + m_\mathrm{int}$. For the latter problem, one can then derive the one-sided power spectral density $S(\omega) \equiv 2 \langle \vert \Tilde{Z}(\omega) \vert ^2 \rangle$, as:
\begin{equation}
\label{eq:PSD}
S(\omega) = 
\frac{2 \langle \vert F_\mathrm{th} \vert ^2 \rangle / (m^{2}\omega_0^4)}{\left[1-\left(\dfrac{\omega}{\omega_0}\right)^2 \right]^2+\left(\dfrac{\omega}{\omega_0 Q} \right)^2} 
= \frac{2k_\mathrm{B}T / (\pi Q m \omega_0^3)}{\left[1-\left(\dfrac{\omega}{\omega_0} \right)^2 \right]^2+\left(\dfrac{\omega}{\omega_0 Q} \right)^2} , 
\end{equation}
where $\langle \cdot \rangle$ denotes the ensemble average, $k_\mathrm{B}T$ is the thermal energy, $\omega_0 = \sqrt{k_{\textrm{c}}/(m_\infty + m_\mathrm{int})}$ is the natural angular frequency, and $Q=(m_\infty + m_\mathrm{int})\omega_0/\gamma$ is the quality factor. The second equality in ~\cref{eq:PSD} is obtained by using the correlator of the noise $\langle F_\mathrm{th}(t)F_\mathrm{th}(t') \rangle = 2 \gamma k_\mathrm{B}T \delta_{\textrm{D}}(t-t')$, where we assumed a white noise through the Dirac distribution $\delta_{\textrm{D}}$ , and where we invoked the fluctuation-dissipation theorem to set the amplitude of the noise.  The experimental power spectral densities are fitted by the function \citep{honig2010lubrication, bowles2011no}:
\begin{equation}
\label{eq:Simplied_PSD}
S(\omega) 
= \frac{c_1}{\left[1-\left(\dfrac{\omega}{\omega_0} \right)^2 \right]^2+\left(\dfrac{\omega}{\omega_0 Q} \right)^2} + c_2, 
\end{equation}
where $\omega_0$ and $Q$ are the key adjustable parameters indicating the position and the width of the resonance, and where $c_1$ and $c_2$ are unimportant extra parameters allowing to accommodate for potential spurious experimental offset and/or prefactor.

\subsection{Power spectral density}
\label{subsection_ExpSpectrum}
\begin{figure}
\centering
\includegraphics[width = 0.95\columnwidth]{{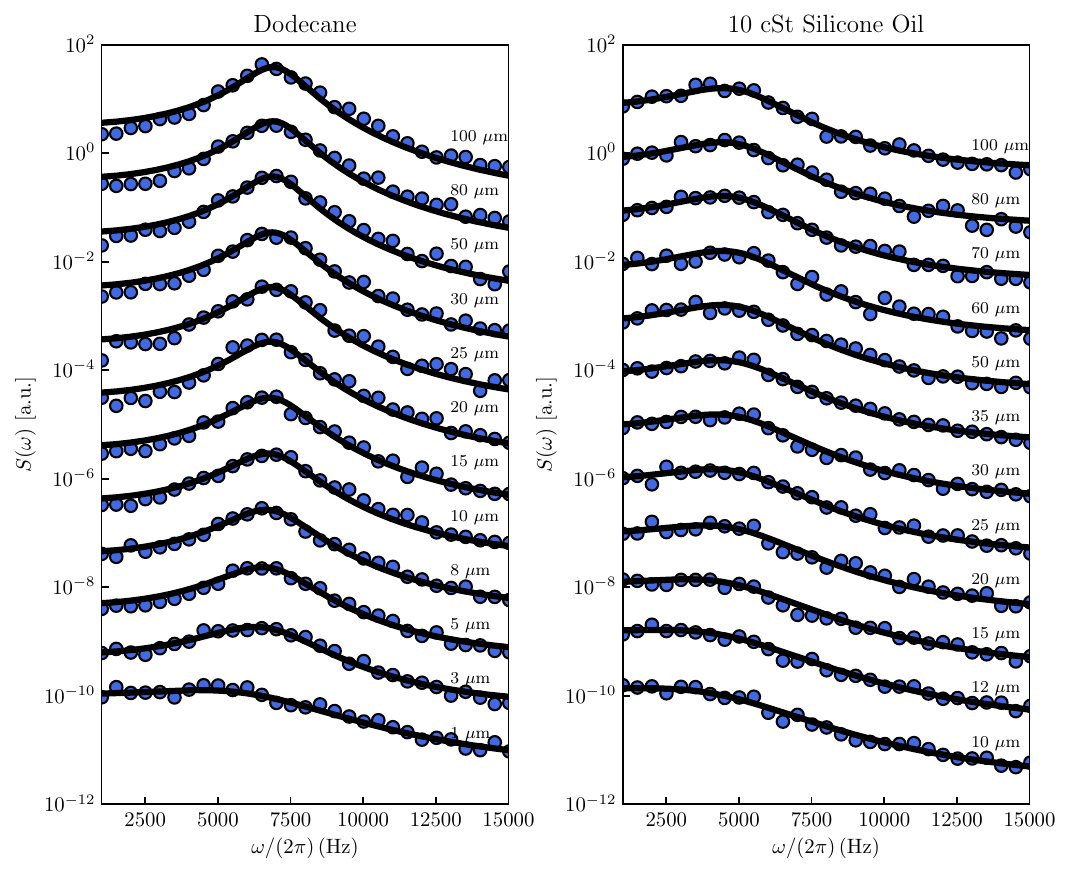}}
\caption{\label{fig:spectrum} Experimental power spectral densities in arbitrary unit [a.u.], for the colloidal probe's vertical position, in dodecane (left) and silicone oil (right), for various probe-wall distances as indicated. The curves are shifted vertically for clarity. The solid lines show the best fits to the damped harmonic oscillator model, using~\cref{eq:Simplied_PSD}. 
}
\end{figure}
\cref{fig:spectrum} displays the power spectral densities for probes immersed in dodecane or silicone oil (water was employed as well, but the similar results are not shown here), and for a variety of probe-wall distances. A well-defined peak can be observed for each spectrum, indicating the fundamental resonance. 
The resonance properties are well described by the damped harmonic oscillator model above. The largest probe-wall distance ($D=100~\mu $m) corresponds to nearly 4 times the sphere radius, so that the hydrodynamic interactions between the probe and the wall can be neglected. At such distances, the bulk resonance frequency $\omega_0^\infty = \sqrt{k_{\textrm{c}}/m_\infty}$ and bulk quality factor $Q_\infty = m_\infty \omega_0^\infty/\gamma_\infty$ are extracted from the fitting procedure, giving respective values of $7070\pm 5$ Hz and $3.3\pm 0.1$ in dodecane and $5320\pm 5 $ Hz and $1.3\pm 0.1$ in silicone oil. In the more viscous fluid (silicone oil), the resonance is broader since the dissipation is larger, as expected. Also, in both liquids, we observe that the resonance is broader as the sphere gets closer to the hard wall, which indicates that the near-wall dissipation is larger as compared to the bulk situation, as expected too. Besides, and interestingly, the natural frequency appears to depend on the viscosity of the ambient fluid, highlighting the fact that the effective mass is not trivial. Moreover, the natural frequency depends on the probe-wall distance. 

\begin{figure}
\centerline{\includegraphics[width=0.95\linewidth]{{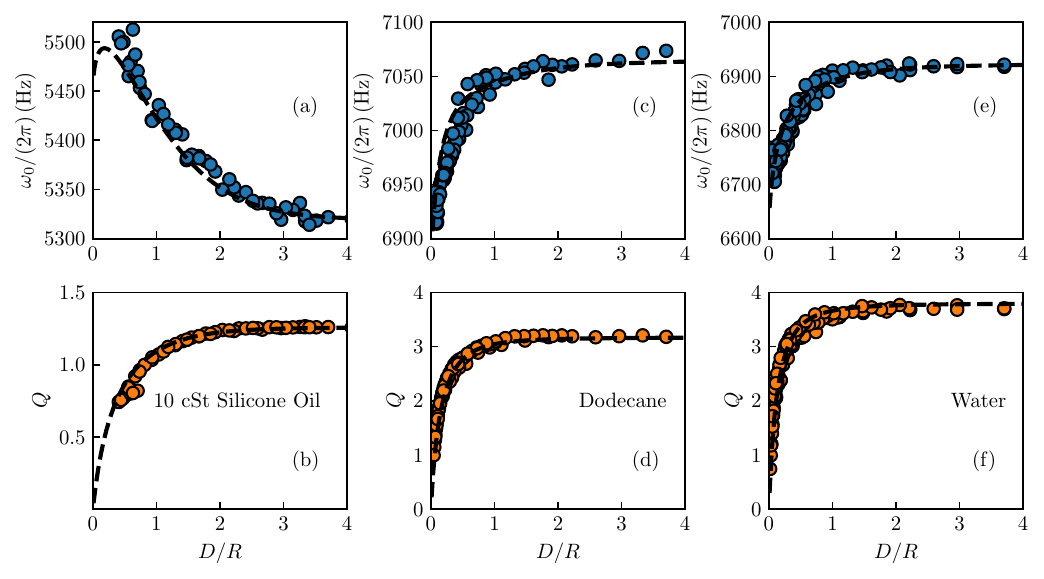}}}
\caption{\label{fig:reuslt_liquid} Natural frequency $\omega_0/(2\pi)$ (blue dots) and quality factor $Q$ (orange dots) of the normal mode of the colloidal AFM probe as a function of dimensionless probe-wall distance $D/R$. The dashed lines represent the natural frequencies and quality factors calculated by \cref{eq:freqencyvariation} and \cref{eq:qualityfactorvariation}, respectively without adjustable parameter. Panels (a,b) show the results for silicone oil ($\eta = 9.3~\mathrm{mPa \cdot s} $, $\rho = 930~\mathrm{kg/m^3}$), with a squared Womersley number of $\wo^2 =2.4$. Panels (c,d) show the results for dodecane ($\eta = 1.34~\mathrm{mPa \cdot s} $, $\rho = 750~\mathrm{kg/m^3}$), with $\wo^2 =18.1$. Panels (e,f) show the results for water ($\eta = 1~\mathrm{mPa \cdot s} $, $\rho = 1000~\mathrm{kg/m^3}$), with $\wo^2 =31.7$.}
\end{figure}

To be quantitative, the fitted values of the natural frequency $\omega_0$ and the quality factor $Q$ are shown in \cref{fig:reuslt_liquid} as functions of the normalized separation distance $D/R$, for the three liquids studied. Intriguingly, we observe an increase of the natural frequency in silicone oil near the wall as compared to the bulk resonance frequency (\cref{fig:reuslt_liquid}(a)), and a corresponding decrease in dodecane (\cref{fig:reuslt_liquid}(c)) and water (\cref{fig:reuslt_liquid}(e)).  
We point out that the probe-wall distances in the present experiments are large enough ($D>0.5~\mathrm{\mu m}$), so that molecular interactions (\textit{e.g.} electrostatic or van der Waals forces) can be safely neglected. Therefore, the changes in natural frequency observed here should only result from hydrodynamic contributions. The following section aims at modeling this intricate behaviour. 

\section{Theory}
\label{sec:model}
\subsection{Governing equations}
\label{subsec:intro_theory}
We aim here at calculating the hydrodynamic force exerted on an immersed sphere moving normally near a rigid, flat and immobile wall. The amplitude of thermal oscillations in the experiments is nanometric, which implies a relatively small Reynolds number for all accessible frequencies. Therefore, we can neglect the convective term of the incompressible Navier-Stokes equations. Nonetheless, the typical resonance frequency is in the kHz range, such that the squared Womersley number $\wo^2 = R^2 \omega / \nu$ is in the 1-50 range. As a consequence, we expect inertial effects to be important. The fluid velocity field $\boldsymbol{v}$ thus satisfies the unsteady incompressible Stokes equations:
\begin{equation}
\label{eq:unsteadyStokes}
\rho \partial_t \boldsymbol{v} = -\boldsymbol{\nabla} p + \eta \boldsymbol{\nabla}^2 \boldsymbol{v},  \quad \quad
 \boldsymbol{\nabla}\cdot \boldsymbol{v} = 0 \, ,
\end{equation}
where $p$ is the hydrodynamic pressure field. Without loss of generality, the sphere's position is supposed to oscillate normally to the substrate at a frequency $\omega$, and with an amplitude $A$, which correspond to a given Fourier mode of the full fluctuation spectrum. Applying the Fourier transform to the  unsteady incompressible Stokes equations, we get: 
\begin{equation}
\label{eq:unsteadyStokes_fourier}
i\rho\omega \tilde{\boldsymbol{v}} = -\boldsymbol{\nabla} \tilde{p} + \eta \boldsymbol{\nabla}^2 \tilde{\boldsymbol{v}},\quad \quad
\boldsymbol{\nabla}\cdot \boldsymbol{\tilde{v}} = 0.
\end{equation}
A no-slip condition is assumed at both the wall and the sphere surfaces, denoted by $\mathcal{S}_{\textrm{w}}$ and $\mathcal{S}_0$ respectively (see \cref{fig:fig1} (b)), leading to the following boundary conditions for the fluid velocity field:
\begin{equation}
\label{eq:boundary-conditions}
    \tilde{\vf}(\boldsymbol{r} \in \mathcal{S}_0) = i\omega A \boldsymbol{e}_z \, , 
    \quad  \quad
    \tilde{\vf}(\boldsymbol{r} \in \mathcal{S}_{\textrm{w}}) = \boldsymbol{0}
    \, ,
\end{equation}
with $\boldsymbol{e}_z$ the unit vector in the $z$-direction. The hydrodynamic drag force applied on the sphere is given by:
\begin{equation}
    \label{eq:def_force} 
    \tilde{\boldsymbol{F}} =  \int_{\mathcal{S}_0} \, \boldsymbol{n} \cdot \tilde{\boldsymbol{\sigma}}\, \mathrm{d}\mathcal{S}_0,
\end{equation}
where $\tilde{\boldsymbol{\sigma}} = -\tilde{p} \mathbf{I} + \eta \left[\boldsymbol{\nabla}\tilde{\vf} + (\boldsymbol{\nabla}\tilde{\vf})^T\right]$ is the fluid stress tensor, and $\boldsymbol{n}$ denotes the unit vector normal to $\mathcal{S}_0$ oriented towards the fluid. To the best of our knowledge, there is no closed-form solution of the problem, in contrast with the steady case (see~\cite{brenner1961slow}). 

By symmetry, the drag force is directed along the $z$ direction, \textit{i.e.} $\tilde{\boldsymbol{F}}=\tilde{F}_z\boldsymbol{e}_z$. Using dimensional analysis, and assuming that the oscillation amplitude $A$ is much smaller than $D$, one can show that the drag force $\tilde{F}_z$ normalized by the bulk Stokes reference $-6i\pi\eta R A \omega$, to form the dimensionless drag force $\tilde{f}_z=\tilde{F}_z/(-6i\pi\eta R A \omega)$, depends only on two dimensionless parameters: i) the Womersley number $\wo$, and ii) the sphere-wall distance relative to the sphere radius $D/R$. As a consequence, the dimensionless hydrodynamic interaction force (see~\cref{subsec:probe-dynamics} and \cref{eq:interaction-force_fourier}) reads:
\begin{equation}
\label{eq:interaction-force_fourier-bis}
    \begin{split}
    \frac{\tilde{F}_\mathrm{int}}{6i\pi\eta R A \omega} &=  \tilde{f}_z\left(D/R \to \infty, \mathrm{Wo}\right) - \tilde{f}_z\left(D/R, \mathrm{Wo}\right)= \frac{(m_\mathrm{int}\omega^2-i\omega \gamma_\mathrm{int})\tilde{Z}}{6i\pi\eta R A \omega}.
    \end{split}
\end{equation} 
Although there is no general analytical solution of~\cref{eq:unsteadyStokes_fourier} with the boundary conditions of \cref{eq:boundary-conditions}, the hydrodynamic drag force has known asymptotic expressions in certain limits, some of which are given in the next two subsections.

\subsection{Large-distance regime}
\label{subsec:asymptotic}
In the infinite-distance limit, the force expression reduces to the BBO equation (see \cref{eq:stokes_isolatedsphere}) for a sphere in an unbounded space, which gives in Fourier space:
\begin{equation}
\label{eq:Basset-fourier}
    \tilde{F}_z = -6i\pi\eta R A\omega \left(1 + \sqrt{-i}\wo   - \frac{i\wo^2}{9}\right), \quad \quad \text{for } \, \, D/R \to \infty.
\end{equation}
The last term of \cref{eq:Basset-fourier} corresponds to an inertial force of added mass $2\pi\rho R^3/3$ and the $\sqrt{-i\mathrm{Wo}}$ term corresponds to the Basset force. The large-distance asymptotic correction to the added-mass contribution due to a rigid wall has been computed using the potential-flow theory, and gives $2\pi\rho R^3 \{1 + 3R^3/[8(R+D)^3]\}/3$ (see~\cite{lamb1932hydrodynamics}). By using a boundary-integral formulation of the unsteady incompressible Stokes equations, \cite{fouxon2018fundamental} have generalized the latter result by including the Basset force, to obtain large-distance the asymptotic drag force, that reads:
\begin{equation}
\label{eq:Fouxon}
\tilde{F}_z = -6i\pi\eta R A\omega\left(1 + \sqrt{-i}\wo - \frac{i\wo^2}{9} + B \frac{R^3}{(D+R)^3}\right), \quad \quad \text{for } \, \, D/R \gg 1,
\end{equation}
where the numerical prefactor $B$ depends on $\wo$ and reads: 
\begin{equation}
    B = \frac{1}{4}\left( 1+\sqrt{-i}\wo-\frac{i\wo^2 }{3} \right) \left[\frac{1}{3}+\frac{3i}{2\wo^2 } \left(1+\sqrt{-i}\wo -\frac{i\wo^2 }{9} \right) \right].
\end{equation}

\subsection{Small-distance regime}
In the limit of small sphere-wall distance, which is of importance for colloidal-probe experiments, the drag force is usually dominated by viscous effects. The out-of-phase component of the force can be described by lubrication theory (see \cite{batchelor1967introduction}), in which the main contribution to the drag comes from the confined region between the sphere and the wall, which leads to the expression: 
\begin{equation}
\label{eq:lubrication}
    \tilde{F}_z = -\frac{6i\pi\eta R^2 A\omega}{D}.
\end{equation}
We stress that the in-phase correction to the latter is still unknown in the lubricated limit. It would be interesting to perform asymptotic-matching calculations on the unsteady Stokes equations (see \cite{cox1967slow}) to obtain a self-consistent expression of the effective added-mass in this limit.

\subsection{Low-Womersley-number regime}
\label{subsec:low_womersley}
As pointed out by \cite{fouxon2018fundamental}, in the small-frequency limit, which corresponds to a small Womersley number, the drag force can be expressed in terms of known integrals, by using the Lorentz reciprocal theorem (see~\cite{masoud2019reciprocal,fouxon2020fluid}). We provide here an alternative derivation of this result. 

We introduce the model steady problem of a sphere moving normally to a surface in a viscous fluid, which corresponds to the problem of \cref{subsec:intro_theory}, at zero frequency, \textit{i.e.}:
\begin{equation}
    \label{eq:def_model_pb}
    \grad \cdot \hat{\boldsymbol{\sigma}} = \boldsymbol{0} \, , 
    \qquad
    \grad \cdot \hat{\boldsymbol{v}} = 0 \, ,
\end{equation}
with the same boundary conditions: 
\begin{equation}
\label{eq:boundary-conditions-model}
    \hat{\boldsymbol{v}}(\boldsymbol{r} \in \mathcal{S}_0) = i\omega A\boldsymbol{e}_z \, , 
    \quad  \quad
    \hat{\boldsymbol{v}}(\boldsymbol{r} \in \mathcal{S}_w) = \boldsymbol{0}
    \, ,
\end{equation}
where $\hat{\boldsymbol{\sigma}}$ and $\hat{\boldsymbol{v}}$ are the fluid stress and velocity fields of the model problem, respectively.
Integrating the Lorentz identity $\nabla \cdot (\tilde{\boldsymbol{\sigma}}\cdot \hat{\boldsymbol{v}} - \hat{\boldsymbol{\sigma}}\cdot \tilde{\boldsymbol{v}}) = i\omega\rho \tilde{\boldsymbol{v}} \cdot \hat{\boldsymbol{v}}$ on the total fluid volume, we obtain:
\begin{equation}
(i\omega A \boldsymbol{e}_z) \cdot \left[\int_{\mathcal{S}_0} \, \hat{\boldsymbol{\sigma}} \cdot \boldsymbol{n}\, \mathrm{d}\mathcal{S}_0 -  \int_{\mathcal{S}_0} \, \tilde{\boldsymbol{\sigma}} \cdot \boldsymbol{n}\, \mathrm{d}\mathcal{S}_0\right] = i\omega\rho \int_\mathcal{V} \tilde{\boldsymbol{v}} \cdot \hat{\boldsymbol{v}} \, \mathrm{d}\mathcal{V},
\end{equation}
where the divergence theorem has been used. Recalling \cref{eq:def_force}, we get: 
\begin{equation}
    \label{eq:force-reciprocal}
    \tilde{F}_z = \hat{F}_z -\frac{\rho}{A} \int_\mathcal{V} \tilde{\boldsymbol{v}} \cdot \hat{\boldsymbol{v}} \, \mathrm{d}\mathcal{V}.
\end{equation}
The force $\hat{F_z}$ and velocity field $\hat{\boldsymbol{v}}$ of the model problem correspond to the ones derived analytically by~\cite{brenner1961slow}, using a modal decomposition. 
The force of the model problem thus reads:
\begin{equation}
\label{eq:force_brenner}
    \dfrac{\hat{F}_z}{6i\pi\eta R \omega A} = \dfrac{4}{3} \sinh(\alpha) \, \sum_{n=1}^\infty \: \dfrac{n(n+1)}{(2n-1)(2n+3)} \, \left\{1-\dfrac{2\sinh[(2n+1)\alpha] + (2n+1)\sinh (2\alpha)}{\left[2\sinh\left((n+\frac{1}{2})\alpha\right)\right]^2 - \left[(2n+1)\sinh(\alpha)\right]^2} \right\},
\end{equation}
with $\cosh(\alpha) = 1 + D/R$. Nevertheless, the unsteady velocity field $\tilde{\boldsymbol{\vf}}$ in \cref{eq:force-reciprocal} is still unknown, so that the drag force $\tilde{F}_z$ cannot be found exactly.

Analytical progress can be made in the low-$\wo$ regime, where the unsteady velocity field can be approximated by the steady solution with $\mathcal{O}(\wo^2)$ corrections, as $\tilde{\boldsymbol{\vf}} = \hat{\boldsymbol{v}} \left[1 + \mathcal{O}(\wo^2)\right]$. In this limit, at leading order in inertial contributions, the drag force reduces to: 
\begin{equation}
    \label{eq:force-reciprocal-lowWo}
    \tilde{F}_z = \hat{F}_z -\frac{\rho}{A} \int_\mathcal{V} \hat{\boldsymbol{v}}^2 \, \mathrm{d}\mathcal{V}.
\end{equation}
The volume integral in ~\cref{eq:force-reciprocal-lowWo} can then be evaluated numerically using the model velocity field provided by \cite{brenner1961slow}.

\subsection{Finite-element method}
\label{subsec:simulations}
\begin{figure}
\centerline{\includegraphics[width=0.95\linewidth]{{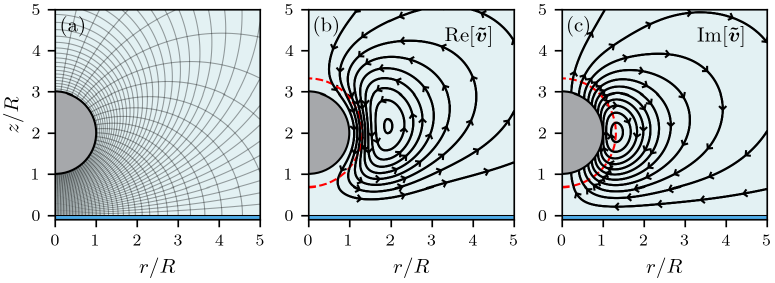}}}	
\caption{\label{fig:numerical} (a) Typical mesh used in the finite-element method. (b) Streamlines of the in-phase flow field, obtained numerically. (c) Streamlines of the out-of-phase flow field, obtained numerically. The squared Womersley number is set to $\wo^2 = 10$, such that $\delta \approx 0.31 R$. The red dashed lines indicate a sphere of radius $R+\delta$.}
\end{figure}

We complement the previous asymptotic expressions of the drag force with full numerical solutions. Using the open-source finite-element library Nutils (see~\cite{nutils7}), we solve \cref{eq:unsteadyStokes_fourier}. The axisymmetric velocity and pressure fields are defined on a $320 \times 320$-element mesh, uniformly spaced on a rectangular domain $[0\leq\tau\leq\alpha,\; 0\leq\sigma\leq\pi]$.
We then use the bipolar coordinate transform: 
\begin{equation}
    r = a \frac{\sin(\sigma)}{ \cosh(\tau) - \cos(\sigma)}, \quad \quad z = a \frac{\sinh(\tau) }{ \cosh(\tau) - \cos(\sigma)},
\end{equation}
with $a = R\sinh \alpha$. The resulting mesh, when axisymmetry is considered, spans the entire domain where $r>0$ and $z>0$, with the exception of a circular region corresponding to the sphere, as shown in~\cref{fig:numerical}. On the symmetry axis ($r=0$), the flow in the radial direction is constrained and the vertical flow is required to be shear-free. At the wall surface ($z=0$), the velocity field is set to zero. Finally, on the surface of the sphere, the radial and vertical velocity components are set to zero and unity (imaginary part) respectively, following \cref{eq:boundary-conditions}. From the calculated velocity and pressure fields, the total force exerted on the particle can be directly computed using~\cref{eq:def_force}. Typical flow fields are shown in \cref{fig:numerical}(b) and (c). 

\section{Results}
\label{sec:results}
\subsection{Drag force}
\begin{figure}
\centerline{\includegraphics[width=0.8\linewidth]{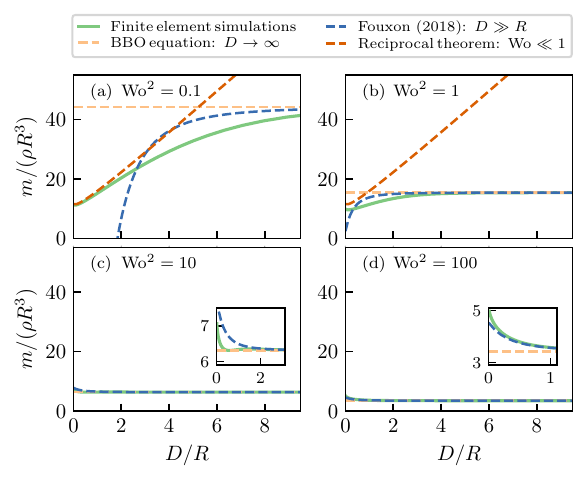}}
\caption{\label{fig:inertial_force} Real part of the total hydrodynamic force, normalized by the inertial force scale, $\mathrm{Re}[\tilde{F}_z]/(\rho R^3 A\omega^2) = m/(\rho R^3)$, as a function of the normalized sphere-wall distance $D/R$. The four panels (a-d) correspond to different Womersley numbers, as indicated. The numerical solutions of section~\ref{subsec:simulations} are shown with solid lines. The bulk Basset–Boussinesq–Oseen force of \cref{eq:Basset-fourier} is displayed with light orange dashed lines. The large-distance asymptotic expression of \cref{eq:Fouxon} is shown with dashed blue lines. The low-$\wo$ expansion of~\cref{eq:force-reciprocal-lowWo} is shown with a dark orange dashed line in panel (a). The insets in panels (c) and (d) show zooms near the wall.}
\end{figure}
\begin{figure}
\centerline{\includegraphics[width=0.8\linewidth]{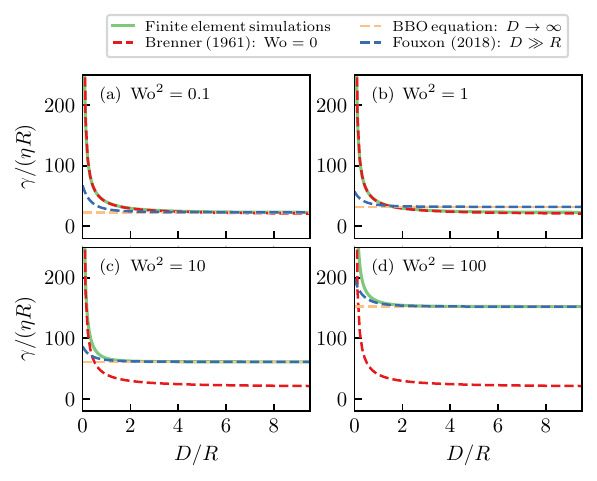}}
\caption{\label{fig:viscous_force} Opposite of the imaginary part of the total hydrodynamic force, normalized by the viscous force scale, $-\mathrm{Im}[\tilde{F}_z]/(\eta R A\omega) = \gamma/(\eta R)$, as a function of the normalized sphere-wall distance $D/R$. The four panels (a-d) correspond to different Womersley numbers, as indicated. The numerical solutions of section~\ref{subsec:simulations} are shown with solid lines. The bulk Basset–Boussinesq–Oseen force of \cref{eq:Basset-fourier} is displayed with light orange dashed lines. The large-distance asymptotic expression of \cref{eq:Fouxon} is shown with dashed blue lines. The viscous solution of~\cref{eq:force_brenner} is shown with red dashed lines.}
\end{figure}

The total hydrodynamic force is decomposed in its in-phase and out-of-phase parts, as $\tilde{F}_z = m A \omega^2 - i \gamma A\omega$, and shown in \cref{fig:inertial_force} and \cref{fig:viscous_force} versus the dimensionless sphere-wall distance. 
First, the Basset-Boussinesq-Oseen force of~\cref{eq:Basset-fourier} agrees well with the simulation results at large distance, for all $\mathrm{Wo}$. The infinite-distance rescaled effective mass is found to increase with decreasing Womersley number as $\sim1/\wo$, for $\wo^2 \ll 1$. This effect arises from the Basset term in ~\cref{eq:Basset-fourier}. Indeed, invoking the velocity scale $A\omega$, one finds a Basset force that scales as $R^2\sqrt{\rho \eta \omega} A\omega \sim \rho R^3 A\omega^2/\wo$. This could rationalize the experimental observations made in ~\cref{fig:spectrum}, where the large-distance natural frequency of the colloidal probe changes in liquids of different viscosities. Conversely, the rescaled damping coefficient increases with increasing Womersley number as $\wo$, for $\wo^2 \gg 1$ (see~\cref{fig:viscous_force}). Here again, this effect originates from the Basset force that also scales as $ \eta R A\omega \wo$.

Interestingly, the behaviour of the rescaled effective mass with dimensionless distance is not universal. For large $\mathrm{Wo}$, the rescaled effective mass decreases with increasing normalized distance. Furthermore, the large-distance asymptotic expression of~\cref{eq:Fouxon} accurately describes the rescaled effective mass in the $\wo^2\gg 1$ regime. Indeed, \cref{eq:Fouxon} is valid as long as the sphere-wall distance exceeds the viscous penetration length, \textit{i.e.} $D\gg \delta = R/\wo$. Near the wall, deviations of the rescaled effective mass from the large-distance asymptotic expression are systematically observed (see insets in~\cref{fig:inertial_force} (c) and (d)), and are comparable to $\sim1$ in magnitude. In sharp contrast, for small $\mathrm{Wo}$, the rescaled effective mass decreases with decreasing dimensionless distance. The typical $\mathrm{Wo}$ value at which the effective-mass variation with distance changes sign is $\wo^2 \approx 5$. In addition, in the small-$\mathrm{Wo}$ regime, the numerical solution agrees well with the asymptotic expression of~\cref{eq:force-reciprocal-lowWo} (see~\cref{fig:inertial_force}(a)) at small dimensionless distances. Eventually, at vanishing sphere-wall distances, the effective mass tends towards a constant value, found numerically to be:
\begin{equation}
    m \approx 11.45 \rho R^3, \quad \quad \mathrm{for}\,\, D \ll R \ll \delta.
\end{equation}
Furthermore, an intermediate regime where the rescaled effective mass increases in an affine manner with the dimensionless distance is observed in Fig.~\ref{fig:inertial_force}(a), as predicted by \cite{fouxon2018fundamental}, as:
\begin{equation}
    m = \frac{9\pi}{4}\rho R^2 (R+D), \quad \quad \mathrm{for}\,\, R \ll D \ll \delta.
\end{equation}
The latter asymptotic expression has been obtained by considering the Lorentz correction to the Stokes drag at large distance (see \cite{lorentz1907allgemeiner,happel1983low}). 

The rescaled damping coefficient decreases with increasing dimensionless distance (see~\cref{fig:viscous_force}). At low $\mathrm{Wo}$, which corresponds to the low-frequency regime, the rescaled damping coefficient is well described at all distances by the steady drag force of~\cref{eq:force_brenner}. However, at large $\mathrm{Wo}$, we observe a transition from the BBO expression at large distance to the steady drag force at small distance. The typical distance at which the transition occurs is  $D \simeq \delta$, which is smaller than $R$. In this regime, the rescaled damping coefficient diverges as $\sim1/D$, as predicted by lubrication theory (see~\cref{eq:lubrication}).

\subsection{Comparison of the model with experiments}
\label{subsec:exp_results}
We now turn to a comparison of the model with experiments. The resonance properties of the colloidal probe are quantified by the natural frequency $\omega_0/(2\pi)$ and quality factor $Q$, as measured by fitting the power spectral density to the harmonic-oscillator model (see~\cref{subsection_ExpSpectrum}). The resulting values of these two quantities were already shown in~\cref{fig:reuslt_liquid}, as functions of the probe-wall distance, for three different liquids of various kinematic viscosities.

Since the natural frequency variations are small, typically on the order of $5\%$ or less of the bulk natural frequency, we perform a Taylor expansion of the natural frequency at first order in $m_{\textrm{int}}/m_{\infty}$: 
\begin{equation}
\label{eq:freqencyvariation}
    \omega_0 = \sqrt{\frac{k_{\textrm{c}}}{m_\infty + m_\mathrm{int}}} \simeq \omega_0^\infty \left(1 - \frac{m_\mathrm{int}}{2m_\infty}\right).
\end{equation}
We then compute the natural frequency at all distances from the numerical simulations, by using ~\cref{eq:interaction-force_fourier-bis}. The Womersley number is set by using the bulk natural frequency, through $\wo^2 = R^2 \omega_0^\infty/\nu$. The resulting $\wo^2$ values are $2.4$, $18.1$ and $31.7$ for silicone oil, dodecane and water, respectively. As shown in~\cref{fig:reuslt_liquid}(a)-(c)-(e), the experimental results agree with the numerical simulation, which confirms that the modification of the natural frequency of the oscillator originates from the hydrodynamic interactions between the sphere and the wall. 

Similarly, we invoke an approximate expression of the quality factor:
\begin{equation}
\label{eq:qualityfactorvariation}
Q = \frac{Q_\infty}{\frac{\omega_0}{\omega_\infty}\left(1+\frac{\gamma_\mathrm{int} Q_\infty \omega_0^\infty}{k_{\textrm{c}}} \right)}\simeq \frac{Q_\infty}{\left(1+\frac{\gamma_\mathrm{int} Q_\infty \omega_0^\infty}{k_{\textrm{c}}} \right)}.
\end{equation}
We then compute the quality factor at all distances from the numerical simulations, by using ~\cref{eq:interaction-force_fourier-bis}, and setting the same $\wo$ values as given above. As shown in~\cref{fig:reuslt_liquid}(b)-(d)-(f), the experimental results agree with the numerical simulation, confirming that the decrease of the quality factor is essentially due to the increase of the viscous Stokes drag as the sphere-wall distance is reduced.

\section{Conclusion}
\label{sec:conclusion}
We investigated the hydrodynamic force exerted on an immersed sphere oscillating normally to a rigid planar wall, by using a combination of colloidal-probe AFM experiments, finite-element simulations and asymptotic calculations. 
The in-phase and out-of-phase components of the hydrodynamic force are obtained from the measurements of the natural frequency and damping of the thermal motion of the probe for various probe-wall distances.
A shift in the natural frequency of the probe was observed with decreasing probe-wall distance, revealing a striking wall-induced unsteady effect: the natural frequency was found to increase with decreasing probe-wall distance in viscous liquids, whereas the opposite trend was observed in low viscosity liquids such as water. By solving the unsteady incompressible Stokes equations numerically, the hydrodynamic force was computed at all distances. The added mass and dissipation increase due to the presence of the wall were then extracted and compared to their experimental counterparts -- with excellent agreement. In addition, at large distance, we recovered the analytical expression derived by~\citet{fouxon2018fundamental}. 
Besides, in the low-Womersley-number limit, the hydrodynamic force could be expressed in a simple integral form using the Lorentz reciprocal theorem, which was validated by the numerical simulations. Beneath the fundamental interest for confined or interfacial fluid dynamics, the present results might be of practical importance for colloidal experiments, because they clarify the hydrodynamic drag acting on a spherical particle near a wall. Essentially, our findings highlight the crucial but overlooked role played by fluid inertia, despite the typically low Reynolds numbers. 

\section*{Acknowledgments}
The authors thank Elie Raphaël, Yacine Amarouchene and Jacco Snoeijer for interesting discussions, as well as Arno Goudeau for preliminary experiments. 

\section*{Funding}
The authors acknowledge financial support from the European Union through the European Research Council under EMetBrown (ERC-CoG-101039103) grant. Views and opinions expressed are however those of the authors only and do not necessarily reflect those of the European Union or the European Research Council. Neither the European Union nor the granting authority can be held responsible for them. The authors also acknowledge financial support from the Agence Nationale de la Recherche under EMetBrown (ANR-21-ERCC-0010-01), Softer (ANR-21-CE06-0029), Fricolas (ANR-21-CE06-0039) and EDDL (ANR-19-CE30-0012) grants, and from the NWO through the VICI Grant No. 680-47-632. They also acknowledge the support from the LIGHT S\&T Graduate Program (PIA3 Investment for the Future Program, ANR-17-EURE-0027). Finally, they thank the Soft Matter Collaborative Research Unit, Frontier Research Center for Advanced Material and Life Science, Faculty of Advanced Life Science at Hokkaido University, Sapporo, Japan.

\section*{Declaration of Interests} The authors report no conflict of interest.

\bibliographystyle{jfm}
\bibliography{jfm}
\end{document}